\documentclass[aps,prl,twocolumn,showpacs,floatfix,superscriptaddress,nofootinbib]{revtex4-1}

\usepackage{amssymb,amsmath,amstext}                
\usepackage{graphicx}                                               
\usepackage{grffile}
\usepackage{epstopdf}                                               
\usepackage{color}                                                     
\usepackage{bm}                                                        
\usepackage{appendix}                                              

\usepackage{ulem}
\usepackage{latexsym}
\usepackage[colorlinks=true,citecolor=blue,linkcolor=magenta]{hyperref}
\usepackage{cancel}
\usepackage{xcolor}
\usepackage{cleveref}
\usepackage{physics}
\usepackage{enumitem}
\usepackage{hyperref}
\usepackage[caption=false]{subfig}

\usepackage{cleveref}

\newcommand*{\icut}{i_{\text{cut}}}
\newcommand*{\lcut}{l_{\text{cut}}}
\newcommand*{\nav}{n_{\text{av}}}

\begin{document} 

\title{Detecting ergodic bubbles at the crossover to many-body localization using neural networks}

\author{Tomasz Szo\l{}dra} 
\affiliation{Instytut Fizyki Teoretycznej, 
Uniwersytet Jagiello\'nski,  \L{}ojasiewicza 11, PL-30-348 Krak\'ow, Poland}
\author{Piotr Sierant} 
\affiliation{The Abdus Salam International Center for Theoretical Physics, Strada Costiera 11, 34151, Trieste, Italy}
\affiliation{Instytut Fizyki Teoretycznej, 
Uniwersytet Jagiello\'nski,  \L{}ojasiewicza 11, PL-30-348 Krak\'ow, Poland}
\author{Korbinian Kottmann} 
\affiliation{ICFO-Institut de Ci\`encies Fot\`oniques, The Barcelona Institute of Science and Technology, Av. Carl Friedrich
Gauss 3, 08860 Castelldefels (Barcelona), Spain}
\author{Maciej Lewenstein} 
\affiliation{ICFO-Institut de Ci\`encies Fot\`oniques, The Barcelona Institute of Science and Technology, Av. Carl Friedrich
Gauss 3, 08860 Castelldefels (Barcelona), Spain}
\affiliation{ICREA, Passeig Lluis Companys 23, 08010 Barcelona, Spain}
\author{Jakub Zakrzewski} 
\affiliation{Instytut Fizyki Teoretycznej, 
Uniwersytet Jagiello\'nski,  \L{}ojasiewicza 11, PL-30-348 Krak\'ow, Poland}
\affiliation{Mark Kac Complex Systems Research Center, Uniwersytet Jagiello{\'n}ski, Krak{\'o}w, Poland}

\date{\today}

\begin{abstract}
The transition between ergodic and many-body localized phases is expected to occur via an avalanche mechanism, in which \emph{ergodic bubbles} that arise due to local fluctuations in system properties thermalize their surroundings leading to delocalization of the system, unless the disorder is sufficiently strong to stop this process. We propose an algorithm based on neural networks that allows to detect the ergodic bubbles using experimentally measurable two-site correlation functions. Investigating time evolution of the system, we observe a logarithmic in time growth of the ergodic bubbles in the MBL regime. The distribution of the size of ergodic bubbles converges during time evolution to an exponentially decaying distribution in the MBL regime, and a power-law distribution with a thermal peak in the critical regime, supporting thus the scenario of delocalization through the avalanche mechanism. Our algorithm permits to pin-point quantitative differences in time evolution of systems with random and quasiperiodic potentials, as well as to identify rare (Griffiths) events. Our results open new pathways in studies of the mechanisms of thermalization of disordered many-body systems and beyond.
\end{abstract}
\date{\today}

\maketitle

\label{sec:intro}

\paragraph*{Introduction.} 
Many-body localization (MBL) \cite{Basko06,Gornyi05} is a phenomenon that prevents strongly disordered quantum many-body systems from reaching thermal equilibrium \cite{Deutsch91, Srednicki94, Alessio16}. In this dynamical phase, the transport is suppressed \cite{Nandkishore15} due to the presence of a complete set of local integrals of motion \cite{Huse14,Ros15,Imbrie16,Mierzejewski18} that account also for the logarithmic growth of entanglement entropy in time evolution \cite{Znidaric08, Bardarson12,Serbyn13b}, or the area-law entanglement of the eigenstates (for recent reviews see \cite{Abanin19,Alet18}). While the properties of MBL phase are considerably well understood, its stability in the thermodynamic limit has recently been vividly debated \cite{Suntajs20e,Panda20,Sierant20b,Kiefer20,Abanin21,Sels20}, and it may depend on various system specific properties \cite{Sierant21}.

The delocalization of MBL was proposed to occur via an ``avalanche'' mechanism \cite{DeRoeck17, Luitz17}. In this approach,
small ergodic regions (``ergodic bubbles'') immersed among insulating blocks delocalize their surroundings and grow at the expense of the localized regions. If the disorder is not sufficiently strong to stop this process, the system becomes delocalized. The avalanche mechanism was soon incorporated into the real space renormalization group approaches \cite{Thiery18,Goremykina19,Dumitrescu19,Morningstar19} that suggest the transition to MBL is of the Kosterlitz-Thouless type. 
This conclusion is supported by a structure of entanglement clusters in eigenstates of system at MBL transition identified in \cite{Herviou19} with the help of quantum mutual information \cite{DeTomasi17}. A related concept of the ``entanglement length'' was introduced in \cite{Gray18}.

The signatures of MBL have been observed experimentally in setups of ultracold atoms 
\cite{Schreiber15,Choi16,Luschen17,Luschen18}, ions \cite{Smith16} or superconducting qubits \cite{Roushan17,Guo21,Gong21}.
Most of the experiments investigated time evolution of initial product states probing correlations between the initial and time-evolved occupations of lattice sites. Strong correlations persisting in the long time limit of time evolution indicate memory of initial state and the onset of MBL \cite{Nicokatz20}. Recent experiments \cite{Rispoli19,Guo20,Morong21}
were able to directly measure the density correlations between various lattice sites, hence opening up new directions in investigations of quantum dynamics.

The aim of this work is to propose a scheme to detect and study the dynamics of the ergodic bubbles at the crossover to MBL relying on the two-site correlation functions that are directly measurable in experiments.
In this way our study parallels recent attempts to understand the delocalization of the dynamics via avalanche mechanism and many-body resonances scenario \cite{Crowley21,Morningstar21,Sels21}. To achieve our goal we employ recurrent neural networks (RNN) that are tailored to time-dependent data and have achieved unprecedented success in natural language processing tasks in recent years \cite{Cho2014,Sutskever2014,Bahdanau2015}.
Our scheme allows us to identify the ergodic bubbles, i.e. the regions of the system, where the dynamics is ergodic, and to study their time evolution. We pin-point the mechanism of delocalization of MBL phase by probing the distribution of the ergodic bubbles. 
Comparing various types of disorder, we detect rare Griffiths regions \cite{Vojta10, Gopalakrishnan15, Gopalakrishnan15a, Agarwal17, Pancotti18} and discuss their impact on delocalization of MBL phase.

\paragraph*{The models.} 
From now on we concentrate on the random-field Heisenberg model with Hamiltonian
\begin{equation}
\label{eq:ham}
H = J\sum_{i=1}^{L} \vec{S}_i \cdot \vec{S}_{i+1} + \sum_{i=1}^L W_i S^z_i,
\end{equation}
 where  $\vec{S}_i$ are spin-1/2 operators, $J=1$ is fixed as the energy unit and
 periodic boundary conditions are assumed. 
 Setting $W_i$ to be independent random variables uniformly distributed in the interval $[-W, W]$,  we obtain 
 the model with random disorder, widely studied in the context of MBL \cite{Santos04a, Oganesyan07, Pal10, Berkelbach10,  Bera15, Enss17, Sierant20, Colmenarez19, Suntajs20,Laflorencie20, Vidmar21}. Estimates of the critical disorder strength vary between $W^{RD}_C\approx 4$ \cite{Luitz15,Mace19b} and $W^{RD}_C\gtrsim5$ \cite{Devakul15, Gray18, Sierant20p}.
 We consider also the quasiperiodic potential $W_i = W\cos\left(2\pi \zeta i + \phi\right)$, where $\phi \in[0,2\pi]$ is a random phase, $W$ characterizes the strength of the potential and the golden ratio $\zeta = \left(\sqrt{5}-1\right)/2$ is used. The transition to MBL phase for that $\zeta$ value occurs at $W^{QP}_C\approx2.5$ \cite{Iyer13, Lee17, Bera17b, Doggen19}. The random potential features rare regions, in which disorder may be anomalously weak or strong. There are no such fluctuations in the quasiperiodic case. Hence, the MBL transition may be qualitatively different in the two cases \cite{Khemani17}.
  
  In what follows we use RNN to analyze the approach to thermal equilibrium in the course of time evolution of system \eqref{eq:ham}.
 The neural networks have been applied to studies of properties of eigenstates of disordered quantum many-body systems  \cite{Schindler17,Hsu18,Huembeli19,Theveniaut19} or in investigations of gross features of their time dynamics \cite{vanNieuwenburg18,Doggen18,Bohrdt20}.

\begin{figure}
	\includegraphics[width=.48\textwidth]{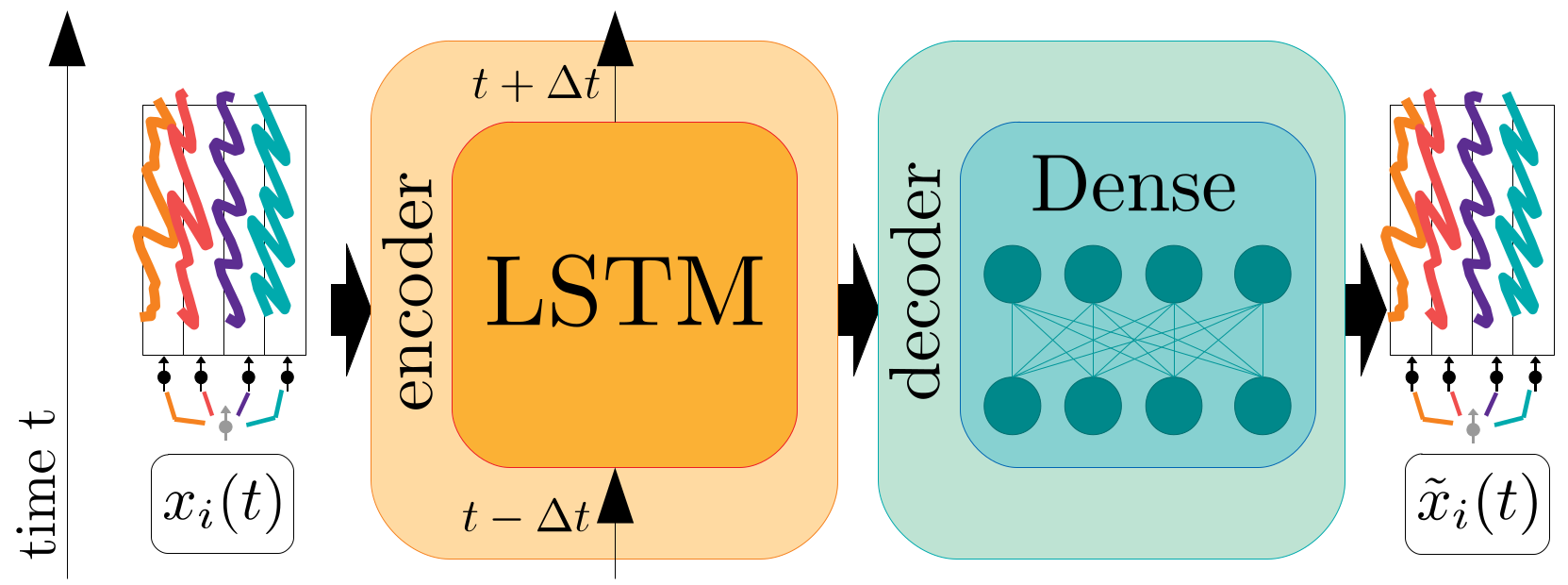}
	\caption{Neural network architecture: A snapshot $x_i(t)$ of two-point correlation functions is input to the multi-channel time series autoencoder. The object $x_i(t)$ is processed by two consecutive LSTM layers in the encoder. The LSTMs additionally receive (pass) a hidden state from (to) the  previous (next) time step, thus creating a time order. The decoder is simply composed of three dense (fully-connected) layers. The network outputs the time series $\tilde{x}_i(t)$ that reconstructs the input signal $x_i(t)$.
  }
	\label{fig:NN_arch}
\end{figure}
\begin{figure}
	\includegraphics[width=\columnwidth]{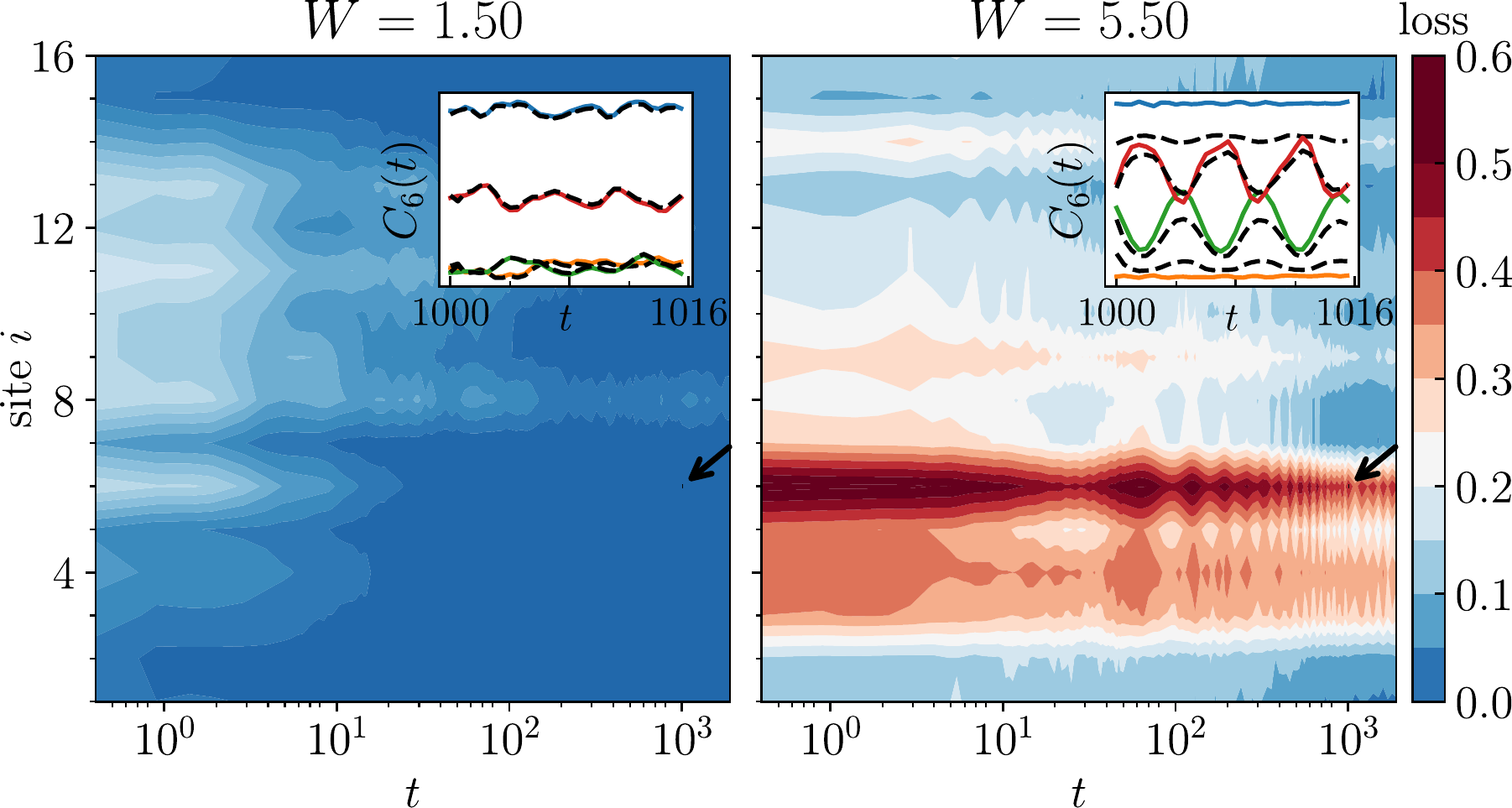}
	\caption{Neural network loss -- a measure of non-ergodicity in the system. In the ergodic regime $W=1.5$, the non-ergodic (high loss) regions quickly disappear; they persist in the MBL phase at $W=5.5$. Both evolutions were calculated for the same uniform disorder realization but with a different amplitude. Insets show the two-site correlations (input of the NN - solid lines), Eq.~\eqref{eq:timeseries}, and the NN output (dashed lines). The NN learned to reconstruct ergodic evolutions perfectly (left inset). It performs significantly worse on the non-ergodic data (right inset) it has not encountered during training, as expected in the anomaly detection scheme.
  }
	\label{fig:single_bubble}
\end{figure}
\paragraph*{Ergodic bubble detection.} We define a set of neighboring spins to be an ergodic \emph{bubble} (or \emph{cluster}) if the local observables associated with them have properties characteristic to an ergodic/thermal phase. In particular, the observables undergo fluctuations in the course of time evolution and this feature may be used for their direct detection. Nevertheless, providing a well-defined criterion for distinguishing the ergodic from the non-ergodic evolutions remains a non-trivial task. In this section we propose an algorithm employing neural networks that learns how the ergodic evolutions ``look like'', and later assigns the time evolutions of experimentally accessible observables to an ergodic or non-ergodic class.

At the core of our framework is the anomaly detection scheme \cite{Pang21} (similar techniques were recently used for mapping out phase diagrams of the quantum many-body systems \cite{Kottmann20a,Kaeming21,Kottmann21b}). Precisely speaking, we consider a neural network of an autoencoder architecture, whose goal is to spot the characteristic features of the input data, ``compress'' it into a latent representation (the \emph{encoder} network) and, from there, to accurately reproduce the original input (the \emph{decoder} network) -- compare FIG.~\ref{fig:NN_arch}. The autoencoder is trained on the \emph{normal} data until it efficiently reproduces it, and then evaluated on a dataset consisting of \emph{normal} and \emph{anomalous} data to identify the \emph{anomalous} data based on the high reconstruction loss. In our case, the \emph{normal} data corresponds to the time series of observables in the ergodic regime at weak disorder, as opposed to the non-ergodic evolutions for large disorder, regarded as the \emph{anomalous} data. 

Specifically, as input data we use the two-site correlation functions
$C_{i,d} (t) = \left\langle S^z_i(t)S^z_{i+d}(t)\right\rangle$.
Let us define a collection of these correlations $x_i(t)$ that is supposed to characterize site $i$ at time $t$. It consists of correlation functions $C_{i,d} (t)$ corresponding to $d_0$ neighbors of site $i$, sampled across $n_t$ discrete points in time
\begin{equation}
x_i(t) = \lbrace C_{i, d}(t + n\Delta t)\rbrace_{\substack{d=-d_0,\dots,-1,1,\dots,d_0\\n=0,\dots,n_t-1}},
\label{eq:timeseries}
\end{equation} 
where, in our implementation we take $\Delta t=0.5$, $d_0=2$ and $n_t=32$. The four-channel time series of length 32 starting at time $t$,  $x_i(t)$, constitutes a single input to the neural network.
Having a trained autoencoder network we can perform an anomaly detection by measuring the reconstruction loss
$   
    l_i(t) = ||\tilde{x}_i(t) - x_i(t)||$, where $||.||$ is the Euclidean norm, see FIG. \ref{fig:single_bubble}.
In order to classify a site $i$ at time $t$, one needs to set a loss cutoff parameter $\lcut$. In our algorithm,
the thermal bubble is defined as a set of adjacent sites which satisfy $l_i \leq \lcut$, that is, for which the network succeeds in reconstructing the temporal evolution of observables with loss no larger than $\lcut$. Thus, we have a quantitative tool to detect the ergodic clusters and to differentiate them from sites at which the anomaly is detected and the dynamics is not ergodic.

Among many neural network architectures, particularly well suited for multichannel time series processing are RNN that naturally utilize the temporal order by recursively including the data from previously processed time steps. In our case, the \emph{encoder} part of the network consists of a special type of RNN with two layers of Long Short-Term Memory (LSTM) architecture \cite{Hochreiter97} with 64 and 32 units, see FIG.~\ref{fig:NN_arch}. As a decoder we use 3 dense time-distributed layers of dimension $128$ and $64$ and $32$. 
The training dataset consists of time evolutions from $t=100$ to $t=2000$ corresponding to $100$ realizations of uniformly distributed disorder of strength $W\in\lbrace0.1, 0.2,\dots 0.5\rbrace$ for system size $L=16$. For a detailed comparison with the networks trained on larger values of disorder, as well as on the quasiperiodic distribution of disorder see \cite{suppl}. 
We stop the training after a certain value $1.7 \cdot 10^{-3}$ of the validation loss is achieved.

\paragraph*{Dynamics of ergodic bubbles.} We denote the average in a given state as $\langle.\rangle_\psi$ and average over 2400 disorder realizations as $\langle . \rangle_W$. To control the risk that a single neural network model accidentally learns a random feature of the data which another model does not, we independently train 30 models on the same data and denote the average over them as $\langle . \rangle_{N}$. All error bars in the following plots correspond to one standard deviation in this averaging. We will consider the distributions of the number of spins $n$ in an ergodic bubble, as well as the average size of ergodic bubbles ${\nav \equiv \langle\langle\langle n \rangle_\psi\rangle_W\rangle_{N}}$. If there are no ergodic bubbles in a time-evolved state $\ket{\psi}$, we manually define $\langle n \rangle_\psi = 0$. Initially, the spins are fully uncorrelated and the initial state is chosen to be a  N\'eel state ${\ket{\psi} = \ket{\downarrow \uparrow \downarrow \uparrow \downarrow \uparrow \dots}}$. The time evolution with Hamiltonian \eqref{eq:ham} is calculated numerically by means of a kernel polynomial method \cite{Tal-Ezer84,Cheby91,Fehske08} for system consisting of $L=24$ spins.
\begin{figure}
	\includegraphics[width=\columnwidth]{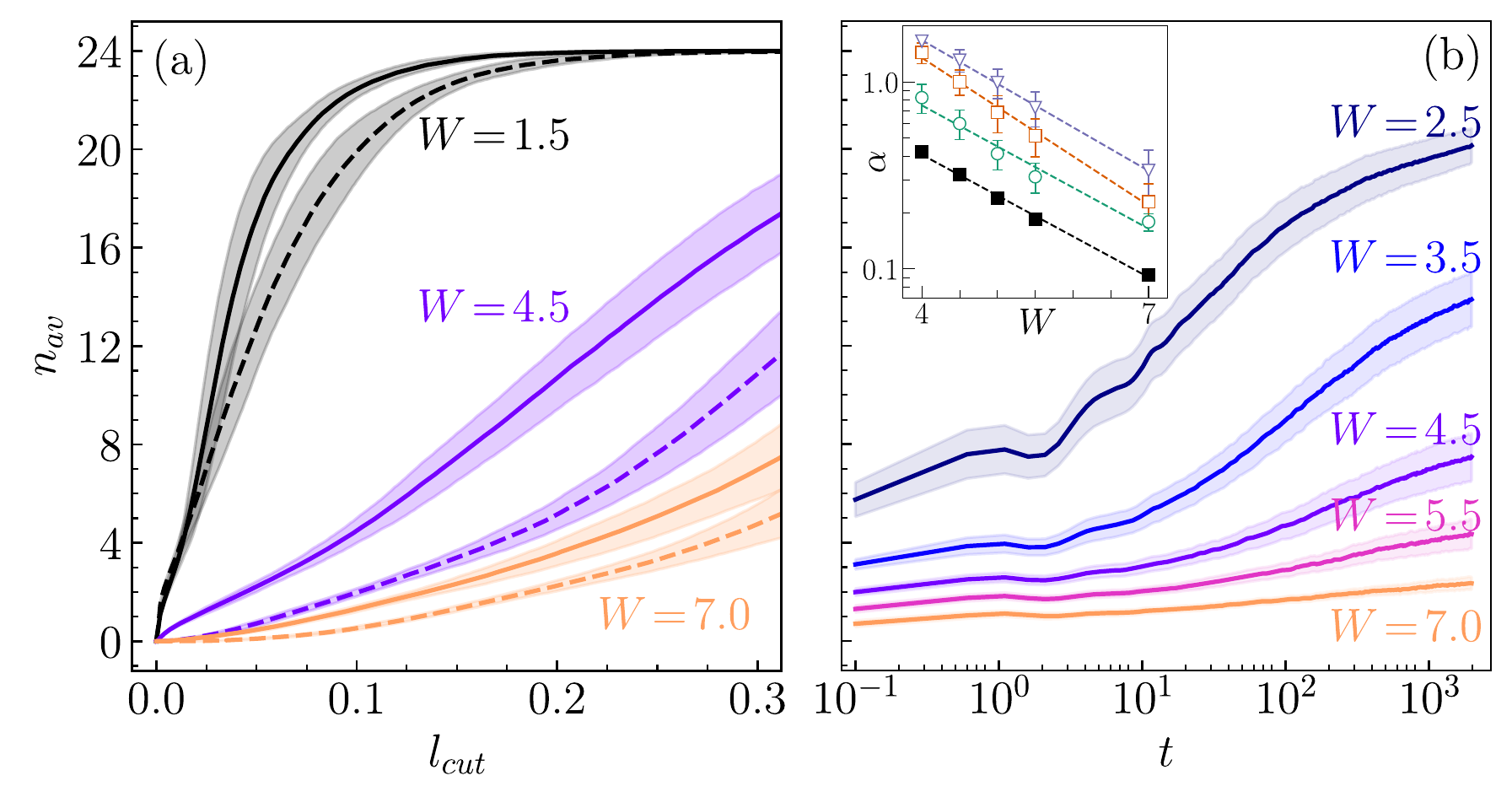}
	\caption{(a) The average size $\nav$ of the ergodic bubbles as a function of the loss cutoff $\lcut$ at time $t=20$ (dashed lines) and $t=1980$ (solid lines) for random disorder. The ergodic ($W=1.5$), critical ($W=4.5$) and MBL ($W=7.0$) regimes differ qualitatively by the dependence of $\nav$ on $\lcut$ and by the amount the bubbles grow in time. 
	(b) Logarithmic in time growth of the average bubble size for $\lcut=0.15$ and several disorder strengths $W$. Inset: see text.} 
	\label{fig:nav}
\end{figure}

We start our studies of the MBL transition by checking how the average size of ergodic bubbles depends on the cutoff $\lcut$, see FIG.~\ref{fig:nav}(a). 
In the ergodic regime the increase of $\nav$ with $\lcut$ is much sharper than for the critical and MBL regimes which signifies that our algorithm indeed detects the prevalent ergodicity in the system.
Moreover, comparing the results for $t=20$ and $t=1980$, we observe that the amount of ergodic regions increases in the course of the evolution, but the growth is much smaller in the MBL than in the ergodic and critical regimes. We may now choose a ``reasonable'' value of the cutoff $\lcut=0.15$, for which, at $W=1.5$, the neural network treats almost the entire chain as a large thermal bubble and in the critical regime it finds a mixture of ergodic bubbles and non-ergodic regions. 

In FIG.~\ref{fig:nav}(b) we present the bubble dynamics for the chosen value of cut-off ${\lcut=0.15}$ and different disorder strengths. Due to a limited temporal resolution of our classification scheme (the time window in which the snapshot $x_i(t)$ of two-site correlation functions has a length of $n_t \Delta t =16$), in the initial stage of the evolution we detect ergodic clusters of non-vanishing size. In the course of time evolution the ergodic bubbles thermalize their surroundings and, consequently, the average size of ergodic bubbles grows with time.
Our results indicate that the growth of the bubbles is logarithmic in time. 

The inset of FIG.~\ref{fig:nav}(b) shows the logarithmic growth rates $\alpha$, fitted from $\nav(t) = \alpha \log t + C$ for different loss cutoffs $\lcut = 0.1, 0.15, 0.2$ (from top to bottom, empty markers), considered as a function of $W$. They are surprisingly well compatible with the similar logarithmic growth rates of the bipartite von Neumann entanglement entropy $\mathcal{S}(\mathcal{A}) = \Tr_{\mathcal{A}} \left(\rho_{\mathcal{A}} \ln \rho_{\mathcal{A}}\right) $
where $\rho_\mathcal{A}$ is the reduced density matrix of subsystem $\mathcal{A}$ consisting of spins indexed by $i=1\dots L/2$ (bottom, full markers). That is, in both cases, $\alpha (W) \approx C \exp(-\beta W)$
where $\beta=(0.55 \pm 0.05)$ while $C$ is different for entropy and clusters for dimensional reasons. 
Importantly, this conclusion holds for arbitrary cut-off $\lcut \in[0.1, 0.2]$,
indicating that our algorithm serves also as an indirect qualitative measure of the entanglement entropy. The decrease of the growth rate of ergodic bubbles with disorder strength $W$ is consistent with the exponential
slow-down of dynamics \cite{Sierant20b, Chanda20m} with increasing $W$ at the crossover to MBL regime.

The qualitative correspondence between the growth of ergodic bubbles and spreading of the entanglement in the system can be explained by a simple argument. 
Let us assume that the chain contains an ergodic cluster of length $n$. 
If the cut that separates the subsystem $\mathcal A$ from its surroundings
does not split the cluster, the contribution from this realization to the average entropy is relatively small. However, if the cut splits the cluster,
the contribution will be much larger because $\mathcal A$ 
contains spins strongly entangled with 
the rest of the ergodic bubble.
According to the volume law of entanglement, this contribution will be proportional to the number of cluster sites included in $\mathcal{A}$ which is proportional to $\nav$. Therefore, we may conclude that $\mathcal{S}$ is proportional to $\nav$. Moreover, we are convinced that the network detects strongly entangled parts of the system 
without the need to directly quantify the entanglement.

\begin{figure}
\includegraphics[width=\columnwidth]{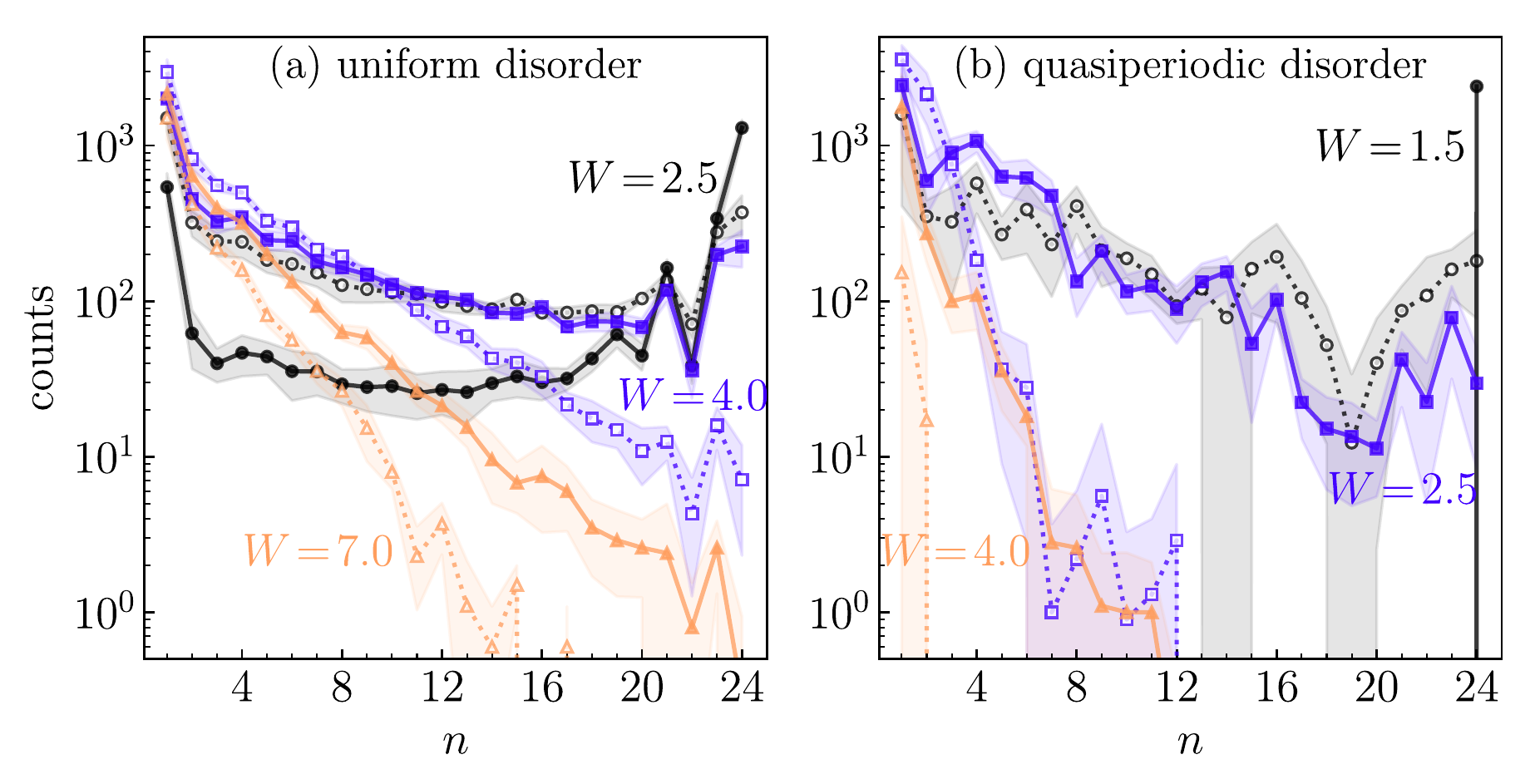}
	\caption{Histograms of the bubble sizes $n$ at time $t=20$ (empty markers) and $t=1980$ (full markers), at a cutoff level $\lcut=0.15$ and (a) uniform and (b) quasiperiodic distribution of disorder, 2400 realizations. }
	\label{fig:hist}
\end{figure}

\paragraph*{Distribution of ergodic bubbles.} 
Further insights into MBL transition are obtained from an analysis of the distributions of the ergodic bubble size $n$. 
In FIG.~\ref{fig:hist}(a) we present such distributions for random disorder, in the initial stage $t=20$, and after a long evolution time $t=1980$. In both cases, in the MBL regime ($W=7$), the distribution decays exponentially as $\exp(-n \gamma )$, consistently with an intuitive expectation that ergodic bubbles correspond to rare events in which the disorder is anomalously weak on $n$ neighboring lattices sites. Indeed, if $p<1$ is the probability to have the weak disorder on a certain site, the probability to find $n$ such sites is $p^n=e^{n \log p}$.
In the vicinity of the critical regime ($W=4$), the ergodic bubbles are distributed, at $t=1980$, according to a heavy-tailed distribution, determination of which is a hard problem given the data changes by less than one order of magnitude. There emerges a thermal peak at $n \approx L$. Indications of a heavy-tailed distribution of the power-law type $n^{-\beta}$ and the presence of thermal peak in the critical regime were observed in renormalization group schemes \cite{Dumitrescu17} as well as in exact diagonalization studies \cite{Yu16}. Exactly at the transition one expects the exponent $\beta$ governing the decay of ergodic clusters to be equal to $2$ \cite{Dumitrescu19}. The smaller value of $\beta$ for $W=4$ suggests that system becomes ergodic in thermodynamic limit at this disorder strength. Here, this value of $\beta$ is obtained around $W=5.3$. While pin-pointing exactly the transition point would require much larger system sizes and evolution times, our results support the picture of the MBL transition occurring via rare avalanches \cite{DeRoeck17}. Finally, on the ergodic side of the crossover ($W=2.5$) we observe abundance of dominant thermal clusters of size $n\approx L$. During time evolution from $t=20$ to $t=1980$, the number of large ergodic bubbles grows in time at the expense of the smaller ones. This process eventually leads to delocalization of the whole system. The transition from an exponential to a heavy-tailed distribution in FIG.~\ref{fig:hist}, with the values of $\beta$ for $W=3.5-7.0$, is presented in \cite{suppl}.

The situation is qualitatively similar for the system with quasiperiodic potential as shown in FIG.~\ref{fig:hist}(b). In the MBL regime ($W=4$) the probability to detect ergodic bubble of size $n$ is exponentially small in $n$. This suggests a presence of similar rare event mechanism as in the case of random disorder. Such a mechanism, in absence of fluctuations in the quasiperiodic potential, might be associated with rare configurations of the state of the system \cite{Gopalakrishnan20}. In the critical regime ($W=2.5$), we find that bubble size is distributed according to a power-law $n^{-2.2}$, although there are no traces of the thermal peak at $n\approx L$. This suggests that delocalization in quasiperiodic system occurs, similarly to the random case, via an avalanche mechanism \cite{Zhang18}. Finally, in the ergodic regime ($W=1.5$), the broad distribution of ergodic bubbles at small time $t=20$ quickly evolves into a single thermal peak at $n\approx L$. The small variance of size of ergodic bubbles both in ergodic and thermal regimes suggests that the MBL transition in quasiperiodic systems is more stable than in the random case \cite{Zhang19}.

\begin{figure}
    \includegraphics[width=\columnwidth]{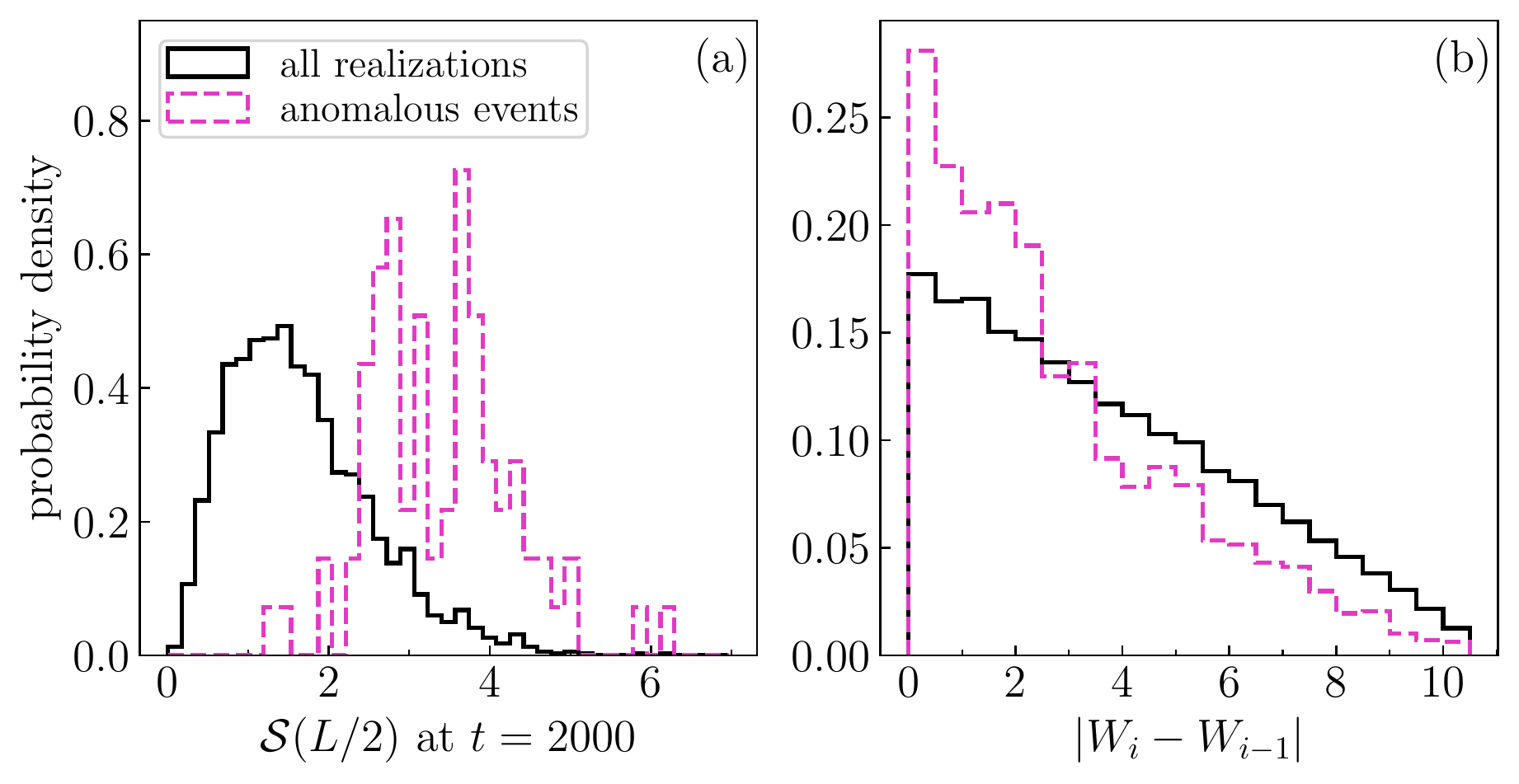}
	\caption{Properties of 2400 realizations of uniform disorder $W=5.5$ (solid black line), including those 74 for which very large clusters $n\geq L-1=23$ were detected in the course of evolution from $t=0$ to $t=1980$ by a single neural network model (dashed pink line). 
	}
	\label{fig:griffiths}
\end{figure}

\paragraph*{Griffiths regions.}
 For the random disorder, even at $W=5.5$, our algorithm detects ergodic bubbles of size $n \geq L-1$, see \cite{suppl} for a corresponding plot similar to FIG.~\ref{fig:single_bubble}. FIG.~\ref{fig:griffiths} shows comparison of properties of all disorder realizations with properties of those anomalous realizations. Exceptionally large bipartite entanglement entropy  $\mathcal{S}(\mathcal{A})$ and strongly resonant values of disorder distinguish these anomalous realizations from the background. This shows that our algorithm detects rare Griffiths events that arise due to fluctuations in the random disorder. For quasiperiodic potential the atypically large bubbles were not detected at all if the cutoffs $\lcut$ were set at the same level. Even if we tune the cutoffs so that the number of detected anomalous realizations is comparable to the uniform disorder case, their features are not distinctive, for details see \cite{suppl}.

\paragraph*{Summary.}
We proposed an algorithm based on neural networks, which allows us to detect and study dynamics of ergodic bubbles in disordered many-body systems -- the subject intensively discussed recently \cite{Crowley21,Morningstar21,Sels21}. The algorithm learns itself features of ergodic time evolution and employs the anomaly detection scheme to identify lattice sites, at which the evolution is nonergodic. Detected ergodic bubbles grow logarithmically in time in the MBL regime, which indicates that our algorithm captures features of entanglement in the system without quantifying it directly. The distributions of the size of ergodic bubble, exponential in the MBL regime and heavy-tailed with thermal peak in the critical regime, support the avalanche scenario of delocalization of MBL phase. Those results are in qualitative agreement with the results obtained when the algorithm of \cite{Herviou19} is used to identify entanglement clusters in state of the system \emph{during time evolution}, as we show in \cite{suppl}. The approach of \cite{Herviou19} relies on the quantum mutual information of numerous subsystems that is very hard to measure in  practice. In contrast, our approach employs only two-site correlation functions readily accessible in present days experiments. Hence, our algorithm allows one for a quantitative investigation of mechanisms of thermalization at MBL transition, not only in the numerics, but also in the experimental setup.

\acknowledgements
\paragraph*{Acknowledgments} 
We would like to thank Fabien Alet for helpful comments and critical reading of the manuscript.
T. S. and J. Z. acknowledge support by PL-Grid Infrastructure and by 
National Science Centre (Poland) under OPUS-18
project  2019/35/B/ST2/00034.
M.L. acknowledges support 
from ERC AdG NOQIA, Agencia Estatal de Investigación (“Severo Ochoa” Center of Excellence CEX2019-000910-S, Plan National FIDEUA PID2019-106901GB-I00/10.13039 / 501100011033, FPI), Fundació Privada Cellex, Fundació Mir-Puig, and from Generalitat de Catalunya (AGAUR Grant No. 2017 SGR 1341, CERCA program, QuantumCAT \_U16-011424 , co-funded by ERDF Operational Program of Catalonia 2014-2020), MINECO-EU QUANTERA MAQS (funded by State Research Agency (AEI) PCI2019-111828-2 / 10.13039/501100011033), EU Horizon 2020 FET-OPEN OPTOLogic (Grant No 899794), and the National Science Centre, Poland-Symfonia Grant No. 2016/20/W/ST4/00314, Marie Sklodowska-Curie grant STRETCH No 101029393.
This project has received funding from the European Union’s Horizon 2020 research and innovation programme under the Marie Skłodowska-Curie grant agreement No 713729 (KK). P.S. acknowledges the support of  Foundation  for
Polish   Science   (FNP)   through   scholarship   START.

%

\section{Supplemental material}

\date{\today}


\maketitle

\setcounter{equation}{0}

\setcounter{figure}{0}

\setcounter{table}{0}

\makeatletter

\renewcommand{\theequation}{S\arabic{equation}}

\renewcommand{\thefigure}{S\arabic{figure}}

\renewcommand{\bibnumfmt}[1]{[S#1]}

\section{Ergodic bubble size histograms}
\begin{figure*}
	\includegraphics[width=\textwidth]{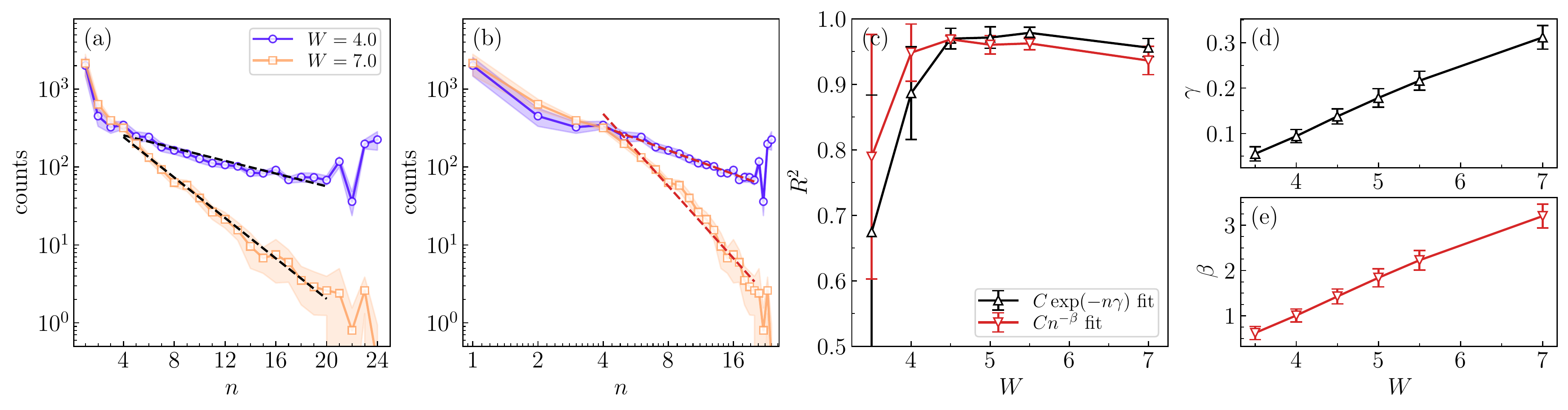}
	\caption{Transition from an exponential to a power-law decay of ergodic bubble size $n$ distributions for loss cutoff $\lcut=0.15$ and system size $L=24$. (a): Bubble size histograms at large time $t=1980$. Exponential fit $C \exp(-n \gamma)$ (dashed line) deviates from the $W=4.0$ datapoints and agrees very well with those for $W=7.0$.(b): same as (a) but with power-law fits $Cn^{-\beta}$ (notice the logarithmic horizontal scale). Here, the fit describes datapoints at $W=4.0$ better than for $W=7.0$. All fits were performed between $n=3$ and $n=20$ only. (c): Measure of the fit quality $R^2 = 1-RSS/TSS$, where $RSS, TSS$ -- residual and total sum of squares, for exponential and power-law fits like in (a) and (b) but for more disorder strengths. Power-law fits the data better in the critical regime $W\lessapprox4.5$. On the MBL side, $W\gtrapprox4.5$, the exponential decay has a lower relative error. (d): Fitted values of $\gamma$. (e): Fitted values of $\beta$.}
	\label{fig:hist_fits}
\end{figure*}

In FIG.~\ref{fig:hist_fits} we show how the ergodic bubble size distribution changes from a heavy-tailed  decay (here we assume a power-law $Cn^{-\beta}$) with a thermal peak in the critical regime $W=4$, into an exponential decay $C \exp(-n \gamma)$ when $W=7$ in the many-body localized phase. All fits are performed on $n=3\dots 22$ to minimize the influence of the thermal peak on the fit, and averaged over 10 independently trained neural networks with the corresponding error bars representing the standard deviation of this averaging. We extract the fitted parameters and present them in FIGs.~\ref{fig:hist_fits}(d),(e). We notice that on the MBL side $\gamma$ increases with an increasing disorder strength. Moreover, it is expected \cite{Dumitrescu19} that at the critical point $\beta=2$ and our results range from $\beta=0.62(14)$ at $W=3.5$ to $\beta=3.20(26)$ at $W=7.0$, with $\beta = 2$ around $W=5.3$.

\begin{figure}
	\centering
	\includegraphics[width=.6\columnwidth]{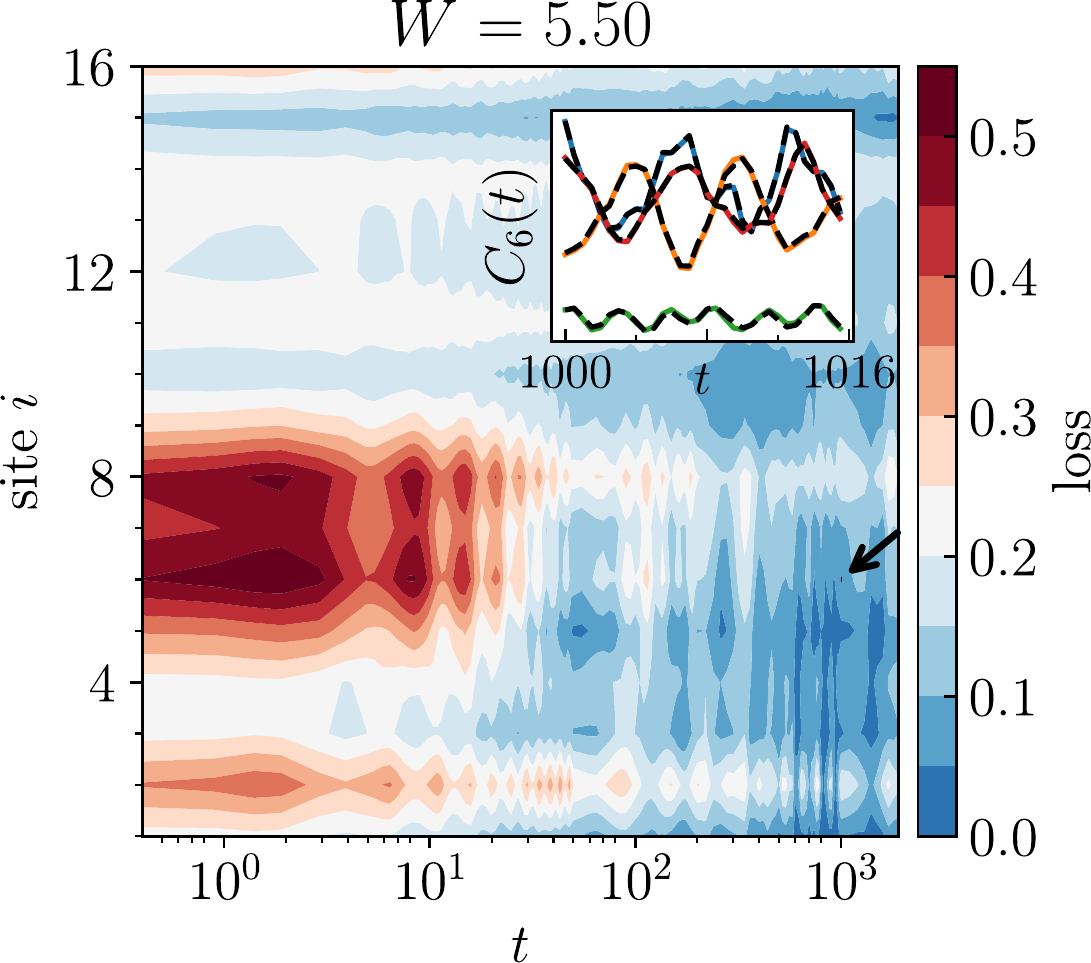}
	\caption{Neural network loss for an anomalous $W=5.5$ disorder realization, detected at the cutoff level $\lcut=0.15$, system size $L=16$. We notice the disappearance of the large loss (non-ergodic) region due to thermalization by its environment.}
	\label{fig:single_bubble_anom}
\end{figure}

In FIG.~\ref{fig:single_bubble_anom} we provide an example of an anomalous realization/Griffiths event, detected by the neural network. An initially strongly non-ergodic regions thermalizes after some time due to the interaction with its ergodic surroundings.
\section{Alternative clustering algorithm based on quantum mutual information}
In this part of the Supplemental material we compare our neural network-based algorithm with a scheme introduced in \cite{Herviou19} that detects so-called \emph{entanglement clusters}. We start with a short revision of this algorithm, point out important differences between entanglement- and ergodic clusters, and then apply this method to the time evolving quantum states, obtaining results in qualitative agreement with our neural network approach.

The input to the entanglement clustering algorithm is the many-body wave function $\ket{\psi}$. After the clusterization procedure, entanglement clusters should contain indices of spins strongly entangled with each other and, at the same time, weakly entangled between separate clusters. The clusters are constructed starting from the whole system and subdividing it recursively into two parts $\mathcal{A},~\mathcal{B}$ containing $n_\mathcal{A},~n_\mathcal{B}$ sites each, at every step minimizing the normalized quantum mutual information (QMI) of the splitting,
\begin{equation}
	0 \leq i(\mathcal{A}, \mathcal{B}) = \frac{\mathcal{S}(\mathcal{A}) + \mathcal{S}(\mathcal{B}) - \mathcal{S}(\mathcal{A} \cup \mathcal{B})}{\min (n_{\mathcal{A}}, n_{\mathcal{B}})} \leq 2 \ln 2,
	\label{eq:qmi}
\end{equation}
where the von Neumann entropy of subsystem $\mathcal{A}$ reads
\begin{equation}
	\mathcal{S}(\mathcal{A}) = \Tr_{\mathcal{A}} \left(\rho_{\mathcal{A}} \ln \rho_{\mathcal{A}}\right),
	\label{eq:S}
\end{equation}
$\rho_\mathcal{A}$ is its reduced density matrix, and periodic boundary conditions are assumed. The number of all such possible bipartitions to choose from grows exponentially with the system size. To make the problem numerically tractable, only a certain class, the \emph{continuous} bipartitions, are left as candidates for the splitting. (``Continuous'' means that the splitting does not introduce holes in a single cluster. For example, a cluster consisting of sites $\{1,2,3,4,5,6\}$ can be divided into $\{1,2,5,6\}$ and $\{3,4\}$ but not into $\{1,3,5,6\}$ and $\{2,4\}$. However, holes in the clusters may occur due to periodic boundary conditions assumed at each step, e.g. splitting $\{1,2,5,6\}$ can result in $\{1,6\}$ and $\{2,5\}$.) Splitting is performed recursively until the cluster size is equal to $1$ and the final representation of the wave function forms a binary tree. Each node of the tree contains indices of the lattice sites, as well as the QMI of the minimal bipartition $i$. Then, the user chooses a certain cutoff of the quantum mutual information $i_{\text{cut}}$ and traverses the tree starting at the root. If the current node contains at least two sites and $i<i_{\text{cut}}$, one goes down to the node's two children. In the opposite case, $i \geq i_{\text{cut}}$, the list of sites at the current node is considered an \emph{entanglement cluster}. A list of all such clusters is the final output of the clustering algorithm at the QMI cutoff level $i_{\text{cut}}$.

There is an important difference in the sense of entanglement and ergodic clusters: every site must be a member of one and only one entanglement cluster whereas it doesn't necessarily have to be assigned to an ergodic cluster; the observables associated with single spins from an entanglement cluster may show either ergodic- or non-ergodic kind of dynamics. This means that the QMI clustering algorithm detects a somewhat different family of clusters than the neural network. On the other hand, it is known that if at least a few spins are strongly entangled, their observables will converge towards thermal values and thermal evolutions, thus large entanglement clusters become equivalent to ergodic clusters. It is, however, not meaningful to compare very small ergodic clusters ($n=1$ or $2$) with similarly small entanglement clusters. Additionally, in our approach, a time series of spin correlations is needed to clusterize the system, whereas in the QMI case the clusters can be defined at time $t$ based on the entanglement properties of the instantaneous wave function $\ket{\psi(t)}$. Furthermore, entanglement clusters can have holes (see the discussion in the preceding paragraph), and NN clusters cannot, but, as argued in ref. \cite{Herviou19}, QMI clusters with holes amount to less than 5-10\% of all clusters in a typical scenario. For a summary of this paragraph see TABLE~\ref{tab:comparison}. 

\begin{figure}
	\includegraphics[width=.93\columnwidth]{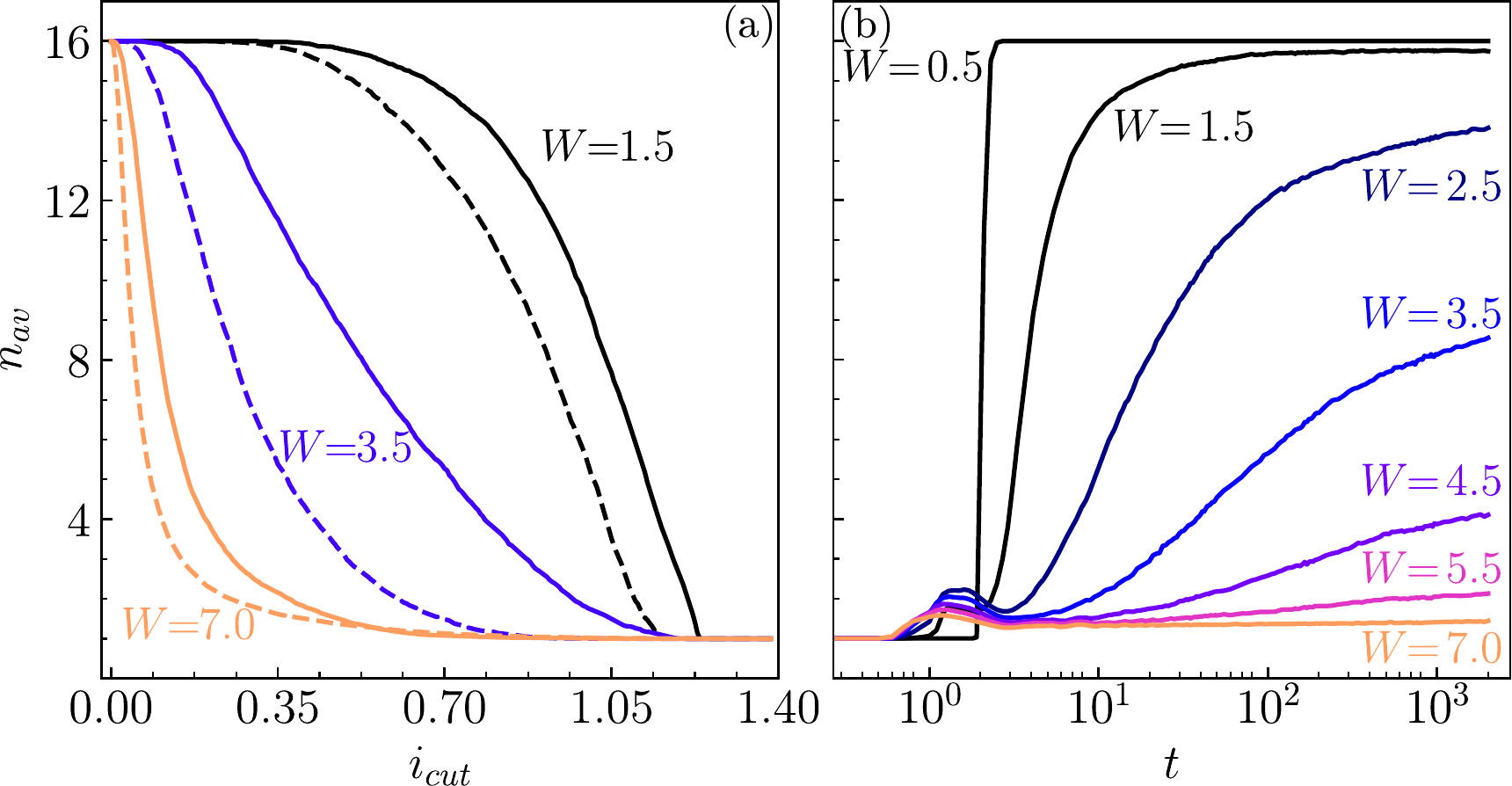}
	\caption{a) Average size of entanglement clusters $\nav$ at time instants $t=20$ (dashed lines) and $t=1900$ (solid lines) for different QMI cutoffs $\icut$ and random disorder strengths $W$, system size $L=16$. b) Average cluster size growth in time for cutoff $\icut=0.5$.}
	\label{fig:growth_QMI}
\end{figure}

\begin{figure}
	\includegraphics[width=\columnwidth]{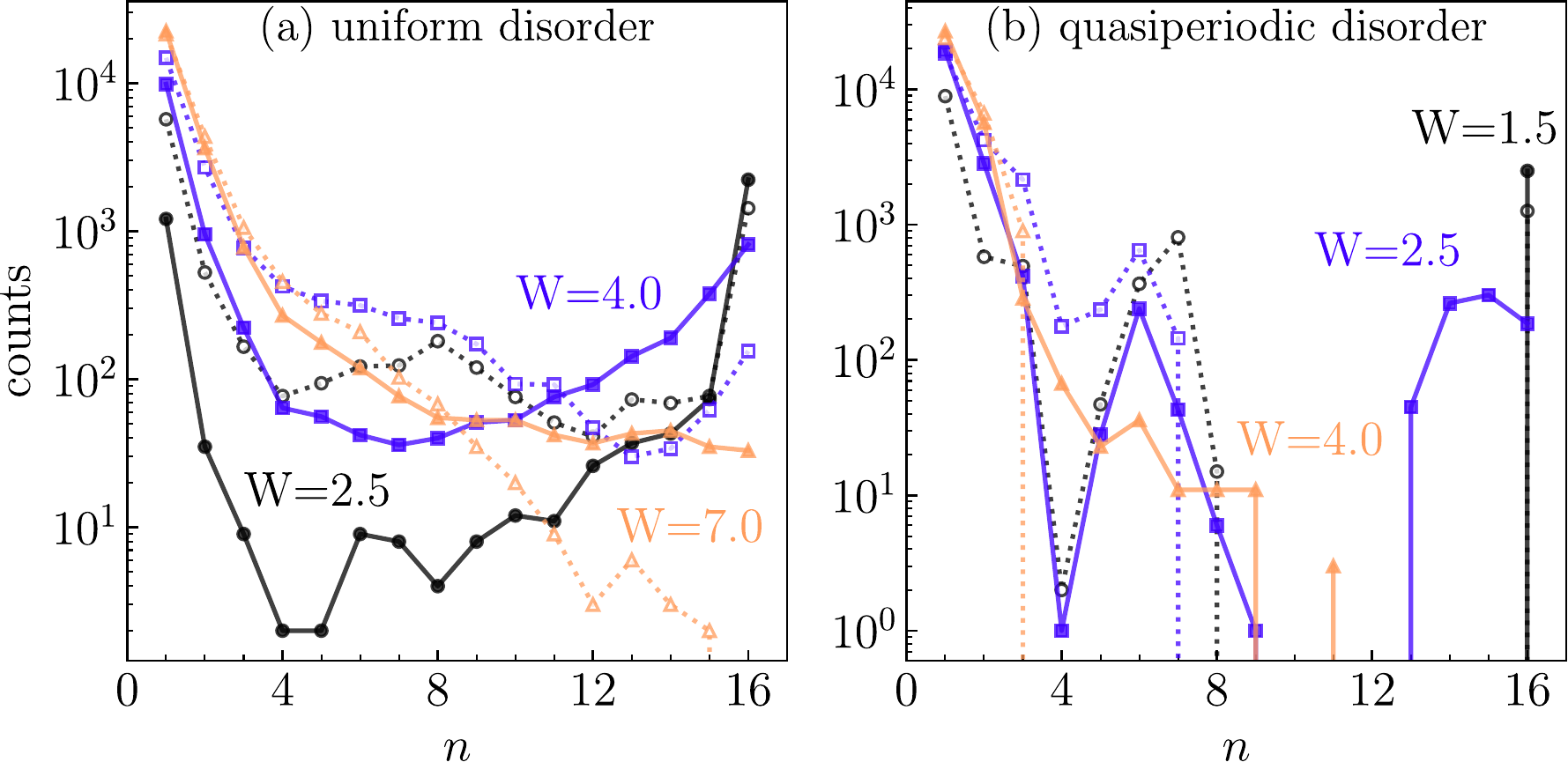}
	\caption{Histograms of the entanglement cluster sizes $n$ detected at the QMI cutoff level $i_{cut}=0.4$ by the QMI clustering algorithm close to the start of the evolution $t=20$ (empty markers) and after a long time $t=1900$ (filled markers) for 2500 disorder realizations with system size $L=16$.}
	\label{fig:hist_growth}
\end{figure}

\begin{table}
	\begin{tabular}{cc}
		Entanglement clusters & Ergodic clusters\\ \hline
		$n\geq 1$ & $n\geq0$ \\ 
		$\sum_i n_i = L$ & $\sum_i n_i \leq L$ \\
		defined for $\ket{\psi(t)}$ & defined for $\lbrace\psi(t), \dots, \ket{\psi(t+n \Delta t)}\rbrace$
	\end{tabular}
	\caption{Comparison of some properties of entanglement clusters and ergodic clusters.}
	\label{tab:comparison}
\end{table}

So far, the QMI clustering procedure has been applied only to the middle-spectrum eigenstates of the disordered Heisenberg Hamiltonian \cite{Herviou19}. Here, we extend those studies to characterize the cluster growth in time and compare the results with the neural network clustering. The system size is limited to $L=16$ due computationally expensive entanglement entropy calculations. As in the Letter, we denote the averaging of the cluster size over a single state as $\langle . \rangle_\psi$ and over disorder realizations as $\langle . \rangle_W$. In FIG.~\ref{fig:growth_QMI}a) we present the average entanglement cluster size $\nav = \langle \langle n \rangle_\psi \rangle_W$ for different cutoff values $\icut$ at time $t=20$ and $t=1900$. We confirm that, similarly to the NN case, the transitions with cutoff are qualitatively different in the ergodic, critical and MBL regimes and that the clusters grow regardless of the cutoff. Next, we select an intermediate value of the cutoff $\icut=0.5$, in FIG.~\ref{fig:growth_QMI}b) we analyze the average cluster size evolution in time and, like for neural networks, observe a logarithmic growth $\nav(t) = \alpha \log t + C$ in the MBL phase. We find no relation between the growth rates $\alpha$ of entanglement entropy and entanglement clusters because the latter strongly depend on the chosen threshold $\lcut$. 

In FIG.~\ref{fig:hist_growth} we plot the histograms of entanglement cluster sizes in the ergodic, critical and MBL regime for uniform and quasiperiodic disorder. In the ergodic case, we notice the disappearance of initially large number of small clusters and the appearance of the power-law distribution {$p(n) \sim n^{-X}$} with a peak for large clusters dominating the distribution after a long time evolution. In the critical regime, after the evolution, we detect a mixture of all cluster sizes. This is in agreement with the main result from the study of eigenstates \cite{Herviou19}, namely that at the critical point small entanglement clusters are entangled together to form larger clusters -- choosing a constant cutoff we are able to observe clusters of nearly all sizes. In contrast, in the localized phase, the clusters become small and independent of each other, with the cluster distribution described by $p(n) \sim \exp(-n\gamma)$ without the thermal peak. All qualitative conclusions about the properties of the cluster distributions are thus in agreement with the NN results presented in the Letter. A one order of magnitude difference in the number of the smallest clusters is connected with the different meanings of ergodic and entanglement clusters, as described in the previous paragraph -- between the two methods one can only compare the properties of the largest clusters.

\begin{figure}
	\includegraphics[width=\columnwidth]{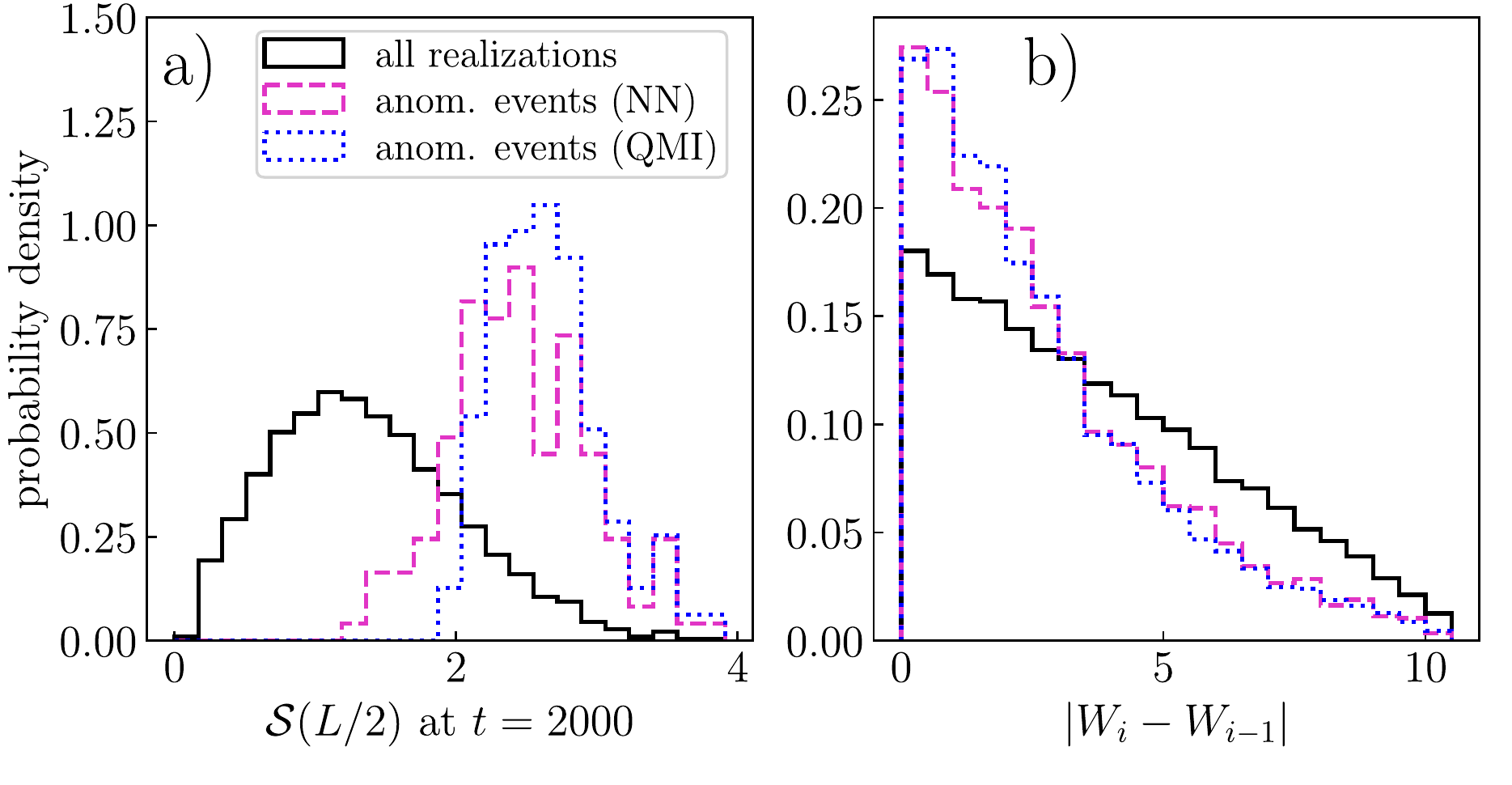}
	\includegraphics[width=\columnwidth]{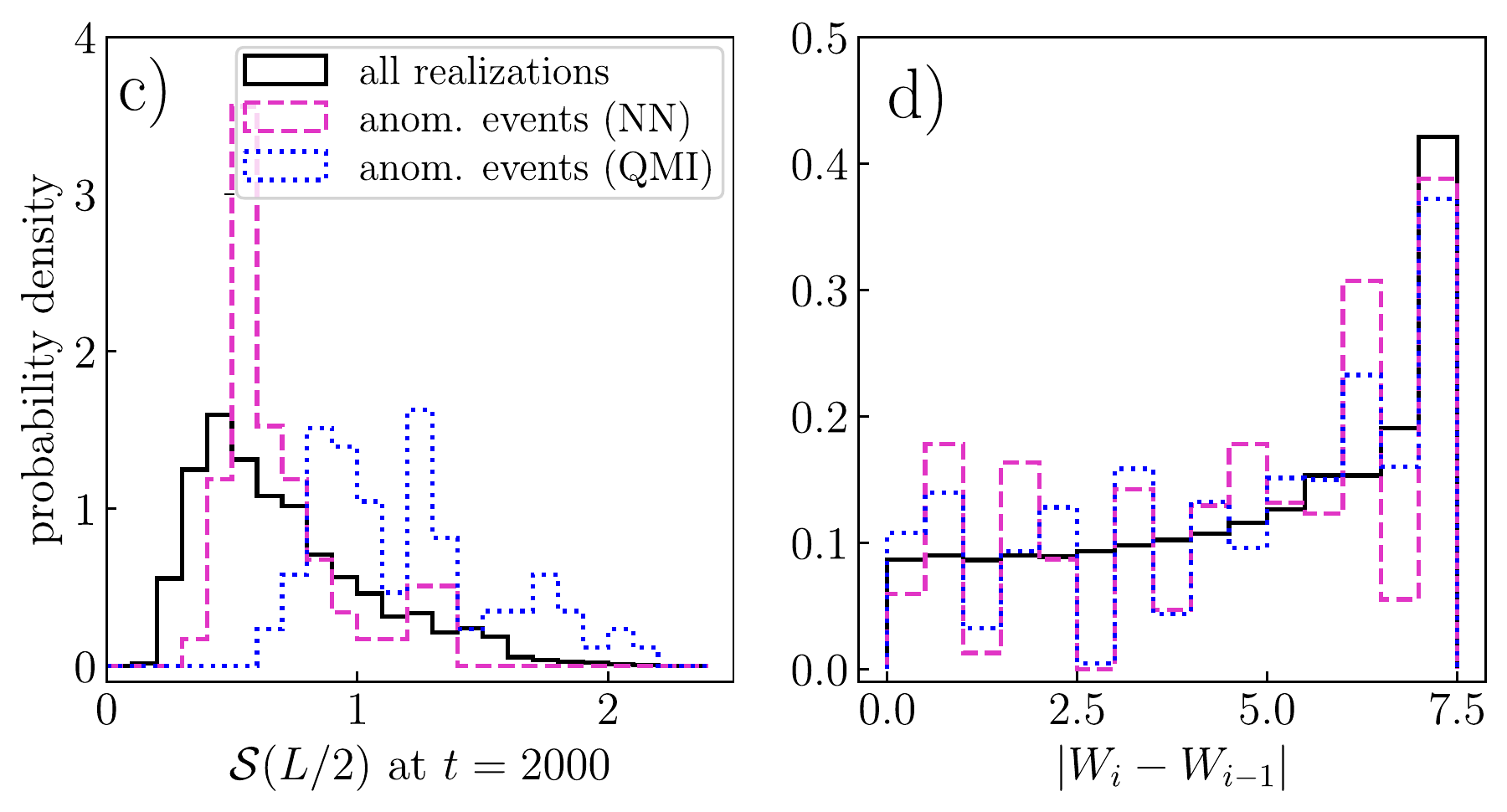}
	\caption{Anomalous realizations for uniform (first row) and quasiperiodic (second row) disorder for system size $L=16$. At the same values of cutoffs as in the uniform disorder case, no anomalous realizations are detected at all. If the cutoffs are tuned to $i_{cut}=0.33$, $l_{cut}=0.09$ (presented here) to obtain a similar number of anomalous realizations as in FIG.~5 from the Letter, these realizations are not as clearly distinguishable from the background as in the uniform disorder case (maybe except for a fraction of the high-entropy states detected by QMI).}
	\label{fig:anomQ}
\end{figure}

\paragraph*{Griffiths regions} We can verify whether the anomalously large clusters, at large disorders, detected by the QMI clustering algorithm, have the same properties as the Griffiths events \cite{Vojta10, Gopalakrishnan15, Gopalakrishnan15a, Agarwal17, Pancotti18} found by the neural network. In FIG.~\ref{fig:anomQ} we plot histograms similar to FIG.~5 from the Letter, except the system size is now $L=16$ (this does not change NN results qualitatively). FIGs.~\ref{fig:anomQ}(a),(b) clearly show, that for the uniform disorder $W=5.5$, the anomalous events detected by both methods at the cutoff levels $\lcut=0.15$, $\icut=0.50$ have nearly the same distributions of entanglement entropies and resonant disorders. Applying both methods to the quasiperiodic disorder with amplitude $W=4.0$ (FIG.~\ref{fig:anomQ}), we do not detect any anomalous realizations at the same cutoff levels. If we tune the cutoffs so that finding a large cluster is more probable ($\lcut =0.25$, $\icut=0.20$), a comparable number of anomalous disorder realizations is detected, but their features are not distinct from the background, except slightly larger entropies found by the entanglement clustering method. This difference is not surprising because entropy properties are the object based on which the entanglement clusters are built.

\section{Neural network training}
In the Letter we present results obtained using a network trained on evolutions corresponding to uniform disorder in the range $W=0.1-0.5$ and time $t=100-2000$, which we assume to be all ergodic. To perform a consistency check and verify how the input data affects the model performance, we also train models on larger values of disorder up to $W=1.0$, as well as on quasiperiodic disorder of the same amplitudes.

\begin{figure}
	\includegraphics[width=\columnwidth]{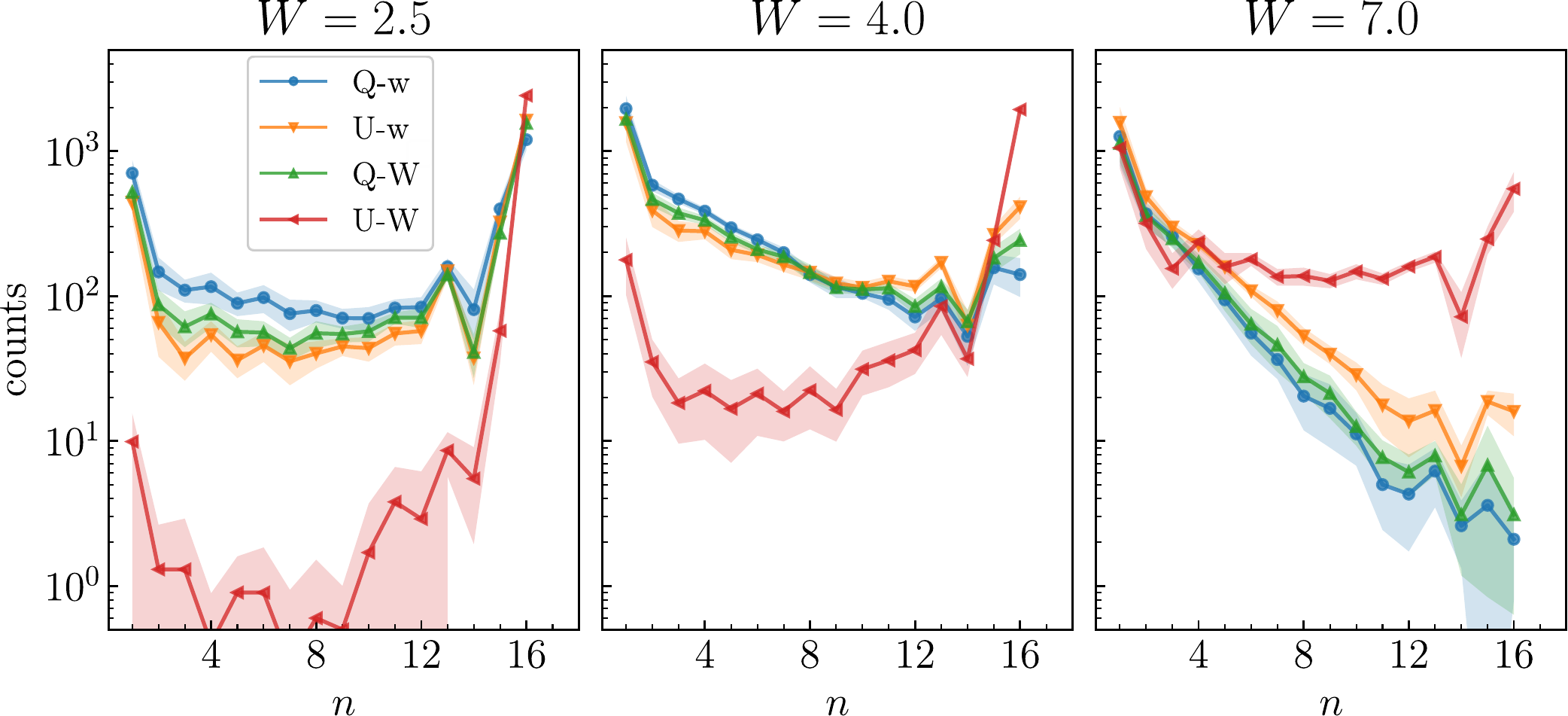}
	\caption{Comparison of ergodic bubble size distributions obtained using neural networks trained on different input data (see TABLE~\ref{tab:robustness}) at loss cutoff $\lcut=0.15$ in the $L=16$ system.}
	\label{fig:NN_comparison}
\end{figure}

\begin{figure}
	\includegraphics[width=\columnwidth]{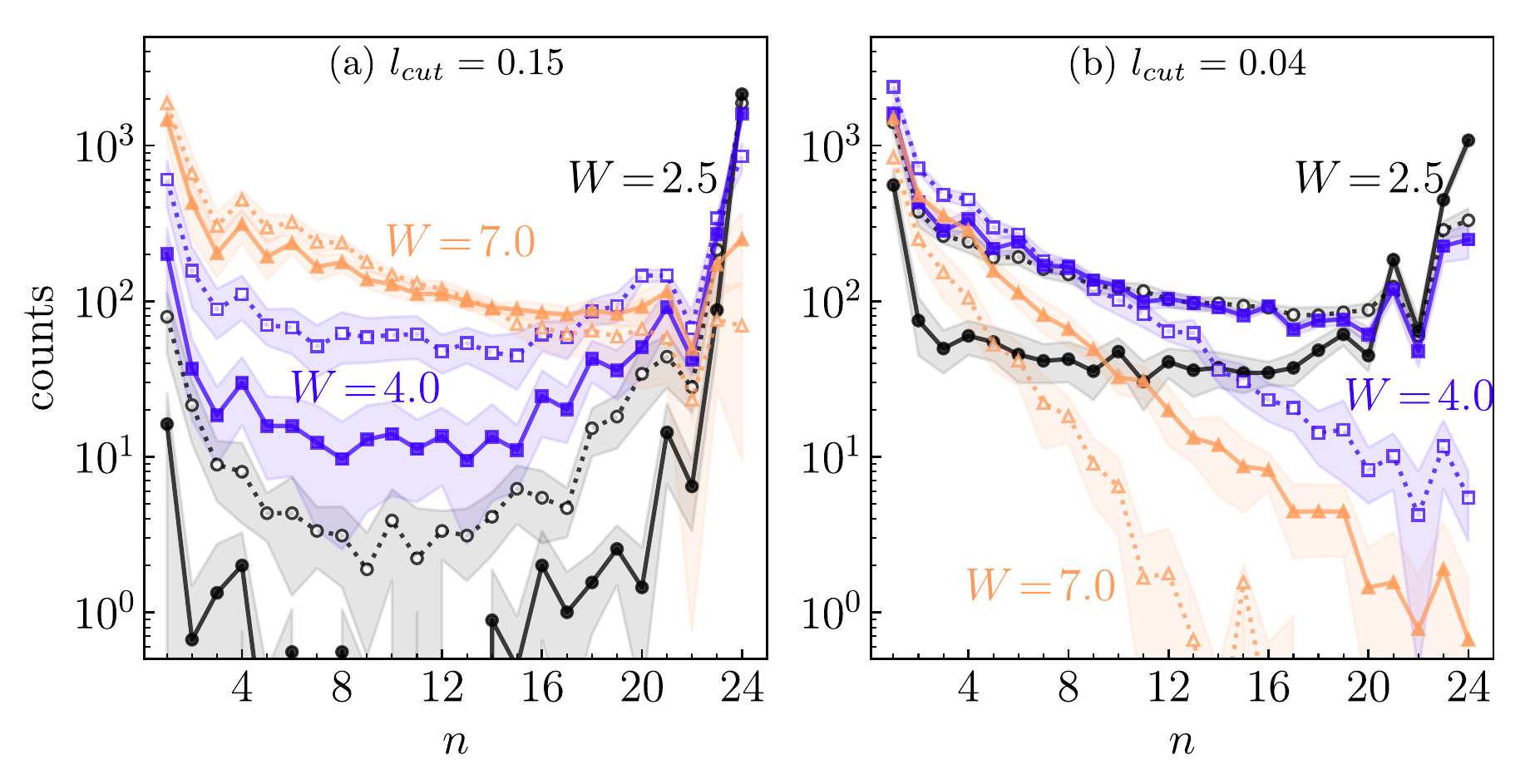}
	\caption{Bubble size histograms for the model trained on $W=0.1-1.0$. For the same value of the loss cutoff $\lcut = 0.15$ as for the model trained on $W=0.1-0.5$ presented in the Letter we obtain a histogram with more large bubbles, even for $W=7.0$. (b) By tuning the cutoff down to $0.04$ we can recover the same qualitative histogram as in FIG. 4a in the Letter. Here, $L=24$.}
	\label{fig:UW_model}
\end{figure}

\begin{table}
	\begin{tabular}{c|c|c}
		model name  & dist. of disorder&  values of disorder  \\ \hline
		"U-w" & uniform & $\lbrace0.1, 0.2,\dots 0.5\rbrace$ \\
		"U-W" & uniform & $\lbrace0.1, 0.2,\dots 1.0\rbrace$ \\
		"Q-w" & quasiperiodic & $\lbrace0.1, 0.2,\dots 0.5\rbrace$ \\
		"Q-W" & quasiperiodic & $\lbrace0.1, 0.2,\dots 1.0\rbrace$ \\
	\end{tabular}
	\caption{Distributions and values of disorder used for training four classes of neural network models. At each value of disorder, time series of its 100 random realizations were used as input. From each model class, 10 neural networks were trained independently on the same data.}
	\label{tab:robustness}
\end{table}

Specifically, the training dataset consists of 100 realizations of disorder from the set called "w": $W\in\lbrace0.1, 0.2,\dots 0.5\rbrace$ or "W": $W\in\lbrace0.1, 0.2,\dots 1.0\rbrace$. We train separately on disorders coming from the uniform (models called "U-w", "U-W") and quasiperiodic ("Q-w", "Q-W") distributions, as shown in TABLE~\ref{tab:robustness}. We stop the training after a certain value $1.7\cdot 10^{-3}$ of the validation loss is achieved. Validation data corresponds to 40 disorder realizations that the network does not use to tune the weights but only to terminate the training when the desired level of loss is achieved. 

In FIG.~\ref{fig:NN_comparison} we present cluster size histograms similar to FIG.~4 from the Letter but for all considered models and only at the end of the evolution $t=1900$ and $L=16$. We observe that all networks except ``U-W'' give consistent results. We suspect that this is due to the fact that in the ``U-W'' training scenario the timescale of the full thermalization of the system is longer than $t=100$.

In FIG. \ref{fig:UW_model} we further evaluate the ``U-W'' networks for $L=24$. They show a similar behavior as for $L=16$ - under the same loss cutoff, they detect large bubbles even at $W=7.0$ (FIG. \ref{fig:UW_model}(a)). However, we find out that upon decreasing the threshold, the histograms from FIG. 4(a) in the Letter can be qualitatively recovered. This means that if one allows for a change in the threshold, the training procedure is robust with respect to the input training data.

It is important to note that the neural networks are trained on the $L=16$ data only. They can later be applied to any system size $L$ which is a strong advantage of the method. It is possible because the two-site correlations the network uses as an input are local observables characterizing a single site and not the whole system of length $L$.

\end{document}